\documentstyle[12pt,epsf]{article}
\newcommand{\vecvar}[1]{\mbox{\boldmath$#1$}}
\begin{document}
hep-ph/0010262

\vspace{0mm}
\begin{flushright}
OU-HEP-364
\end{flushright}
\vspace{-8mm}
\begin{center}
\large{\bf Light flavor sea-quark asymmetry for the
spin-dependent distribution functions of the
nucleon}\footnote{Invited talk at the XIV International Seminar
on High Energy Physics Problems ``Relativistic Nuclear Physics
and Quantum Chromodynamics'', Dubna, 24-29 September, 2000}

\vskip 4mm
M.~Wakamatsu$^{\dag}$

\vskip 4mm

{\small

{\it Department of Physics, Faculty of Science,}\\
{\it Osaka University, Toyonaka, Osaka 560, Japan}
\\
$\dag$ {\it
E-mail: wakamatu@miho.rcnp.osaka-u.ac.jp}
}
\end{center}

\vskip 1mm

\begin{center}
\begin{minipage}{150mm}
\centerline{\bf Abstract}
\begin{small}
\ \ \ The theoretical predictions of the chiral quark soliton
model for the unpolarized and longitudinally polarized structure
functions of the nucleon are compared with recent high energy
data. The theory is shown to explain all the qualitatively
noticeable features of the existing experiments, including the
light flavor sea-quark asymmetry for the unpolarized distribution
functions established by the NMC measurement as well as very
small quark spin fraction of the nucleon indicated by the EMC
measurement. Another unique feature of the model is that it
predicts sizably large isospin asymmetry also for the spin-dependent
sea-quark distribution functions.
\end{small}
\\
\vskip -3mm
\begin{small}
Nucleon Spin Contents, Chiral Soliton, Large $N_c$ QCD
\end{small}
\end{minipage}
\end{center}

\vspace{2mm}
\noindent
\begin{large}
{\bf 1. Introduction}
\end{large}
\vspace{3mm}

\ \ \ Undoubtedly, the EMC measurement in 1988
is one of the most surprising discoveries in the recent
studies of nucleon structure functions \cite{EMC88}.
In my opinion, also very important is the NMC measurement \cite{NMC91},
which has established the flavor asymmetry of sea-quark
distributions in the nucleon. Why is the 2nd observation so important?
Because it is the first clear manifestation of nonperturbative
QCD dynamics in high-energy deep-inelastic scattering observables.
In fact, the most popular explanation of the NMC measurement is
due to the pion cloud effect, which is an inevitable physical
consequence of Nambu-Goldstone realization of chiral symmetry.
 
By now, the NMC observation is known to be explained by several chiral
models including the intuitively simplest model, i.e. the Pion 
Cloud Convolution Model \cite{Kumano98} and also the CQSM which we are
advocating.
However, the physical contents of these two models are not necessarily
the same. In fact, we already know that an important advantage
of the CQSM over the former is that it is able to explain unexpectedly
small quark spin contents of the nucleon simultaneously with the above
sea-quark asymmetry \cite{WY91}.
Another important difference will be revealed
by considering the following natural question :
"Do we expect flavor asymmetric sea also for spin 
dependent PDF?" As we shall discuss below, the answers of the above 
two models seem quite different.
What makes this difference is an interesting theoretical question,
which I shall address in my present talk.

To understand the whole story, it would be helpful to
remember some basic features of the CQSM \cite{DPP88},\cite{WY91} :

\begin{itemize}

\item First of all, it is a relativistic field theoretical model of
baryons effectively incorporating the basic ingredients of large
$N_C$ QCD.

\item At large $N_c$, a nucleon is thought to be an aggregate of
$N_c$ valence quarks and infinitely many Dirac sea quarks bound by
self-consistent pion field of hedgehog shape.

\item Canonically quantizing the spontaneous
rotational motion of the symmetry breaking mean field configuration,
we can perform nonperturbative evaluation of any nucleon observables 
with full inclusion of valence and Dirac sea quarks.

\item Finally, but most importantly, {\it only 1 parameter} of the model
(that is the dyanamical quark mass $M$) was already fixed by low energy
phenomenology, so that we can give {\it parameter-free predictions}
for the parton distribution function at the low renormalization scale.

\end{itemize}

\vspace{6mm}
\noindent
\begin{large}
{\bf 2. CQSM and Twist-2 quark distribution functions}
\end{large}
\vspace{3mm}

  We start with the standard theoretical definition of the quark 
distribution functions.
\begin{eqnarray}
   q (x) \ \ &=& \ \ \frac{1}{4 \pi} \int^{\infty}_{-\infty} 
   d z^0 \,\,e^{\,i \,x \,M_N \,z^0} \nonumber \\
   &\times& \langle N (\vecvar{P}=0) \,| \,
   \psi^\dagger (0) \,O \,
   \psi(z) \,| \,N 
   (\vecvar{P}=0) \rangle \,\,
   |_{z^3 = -z^0,\, z_{\perp} = 0} \,.
\end{eqnarray}
The following novel $N_c$ dependencies follow from the theoretical
structure of the model, i.e. the mean-field approximation and the 
subsequent perturbative treatment of collective rotational
motion \cite{DPPPW96}--\cite{WK99} :
\begin{eqnarray}
  u(x) + d(x) \ &\sim& \ N_c \,\,
  [ \,O(\Omega^0) \ + \ \ \ 0 \ \ \,] \ \ \ \ \sim \ \  O(N_c^1) \, ,\\
 u(x) - d(x) \ &\sim& \ N_c \,\,
 [ \, \ \ 0 \ \ \ + \ O(\Omega^1) \,] \ \ \ \ \sim \ \ O(N_c^0) \, ,\\
 \Delta u(x) + \Delta d(x) \ &\sim& \ N_c \,\,
 [ \, \ \ 0 \ \ \ + \ O(\Omega^1) \,] \ \ \ \ \sim \ \ O(N_c^0) \, ,\\
 \Delta u(x) - \Delta d(x) \ &\sim& \ N_c \,\,
 [ \,O(\Omega^0) \ + \ O(\Omega^1) \,]
 \ \ \sim \ \ O(N_c^1) \ + \ O(N_c^0) \, .
\end{eqnarray}
Notice that, because of the peculiar spin-isospin correlation imbedded
in the hedgehog mean field, there is no leading-order $N_c$ contribution
to the isovector unpolarized distribution  as well as to the
isoscalar longitudinally polarized one. This especially 
means that the isoscalar (or flavor singlet) axial charge is
parametrically smaller than the isovector one, in conformity with
the EMC observation.

\vspace{6mm}
\noindent
\begin{large}
{\bf 3. Theory versus Experiments}
\end{large}
\vspace{3mm}

To make a comparison with high energy data, we take the predictions 
of the CQSM as initial distributions given at low energy scale.
Here, we assume that the distribution functions of the $s$-quark and 
gluons are both zero at this low energy scale.
The scale dependencies of the distribution functions are taken into
account by using Fortran code of the standard evolution equation at the
NLO \cite{HKM98}.
The starting energy of this evolution is fixed to be 
$Q^2 = 0.30 \,\mbox{GeV}^2$.
The model predictions for the twist-2 distribution functions are
summarized in Fig.1.
In this figure, the long-dashed and dash-dotted curves stand for 
the contribution of $N_c$ valence quarks and of Dirac-sea quarks,
respectively, while their sums are denoted by solid curves.
The functions in the negative $x$ region should be interpreted as 
antiquark distributions according the rule :
\begin{eqnarray}
 u(-x) \ \pm \ d(-x) \ \ &=& - \,
 [\,\bar{u}(x) \ \pm \bar{d}(x) \,]
 \hspace{15mm} (0 < x < 1) \, , \\
 \Delta u(-x) \ \pm \ \Delta d(-x) &=& \Delta \bar{u}(x) \ \pm \ 
 \Delta \bar{d}(x) \hspace{14mm} (0 < x < 1) \, .
\end{eqnarray}
The crucial role of the Dirac-sea contribution is most clearly seen
in the isoscalar unpolarized distribution.
Here, the ``valence-quark-only'' approximation leads to positive
$u(x) + d(x)$ in the negative $x$ region, thereby violating the
positivity of the antiquark distribution \cite{DPPPW96}.
  The effect of Dirac-sea quarks is very important also for the isovector
unpolarized distribution function. Especially interesting here is the 
fact that $u (x) - d (x) > 0$ in the negative $x$ region, which means that 
$\bar{u} (x) - \bar{d} (x) < 0$, in conformity with the NMC
observation \cite{WK99},\cite{PPGWW99}.

Turning to the longitudinally polarized distributions, one observes
very different $x$ dependencies between the isoscalar and isovector ones.
One interesting feature of the isoscalar distribution is its sign charge
in the small $x$ region. We shall later see that this sign change is 
just what is required by the recent deuteron data \cite{WW00A}.      
Turning to the isovector distribution, we notice that the effect of 
Dirac-sea quarks has a peak of positive sign around $x \simeq 0$.
Noteworthy here is the positivity in the negative $x$ region. 
It means that sea-quark or anti-quark distribution breaks isospin
SU(2) symmetry also for the longitudinally polarized
distributions \cite{WW00A}.
\begin{figure}[thbp]
\begin{center}
\epsfxsize=12.0cm\leavevmode\epsfbox{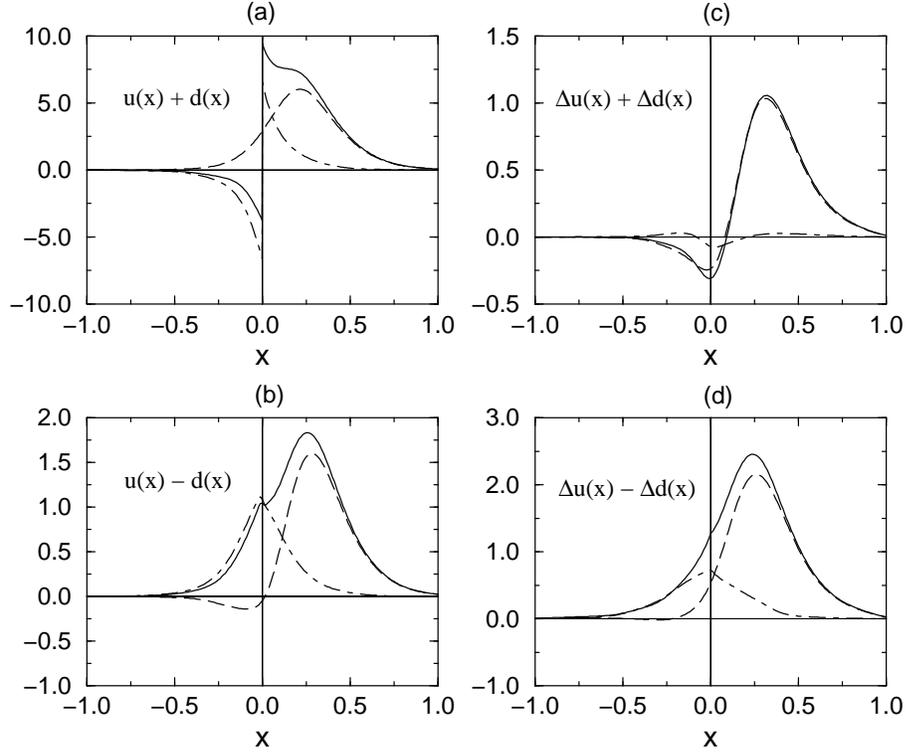}
\end{center}
\renewcommand{\baselinestretch}{1.00}
\caption{\small{The theoretical predictions of the CQSM for the unpolarized
distributions $u(x) + d(x)$ and $u(x) - d(x)$ as well as for the
longitudinally polarized distributions $\Delta u(x) + \Delta d(x)$
and $\Delta u(x) - \Delta d(x)$. In all the figures, the long-dashed
and dash-dotted curves respectively stand for the contributions of
the discrete valence level and that of the Dirac continuum in the
self-consistent hedgehog background, whereas their sums are shown
by the solid curves.}} 
\renewcommand{\baselinestretch}{1.00}
\end{figure}
\begin{figure}[htbp]
\begin{center}
\epsfxsize=11.0cm\epsfysize=7.0cm\leavevmode\epsfbox{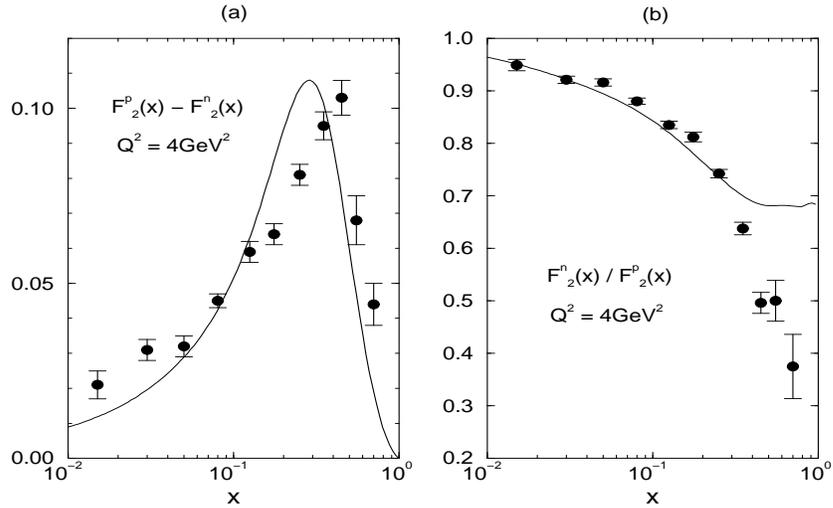}
\end{center}
\renewcommand{\baselinestretch}{1.00}
\caption{\small{The predictions for $F^p_2 (x) - F^n_2 (x)$ and
$F^n_2 (x) / F^p_2 (x)$ at $Q^2 = 4 \,GeV^2$ are compared with the
NMC data given at the corresponding energy scale.}}
\renewcommand{\baselinestretch}{1.00}
\end{figure}
\begin{figure}[hbtp]
\begin{center}
\epsfxsize=12.0cm\leavevmode\epsfbox{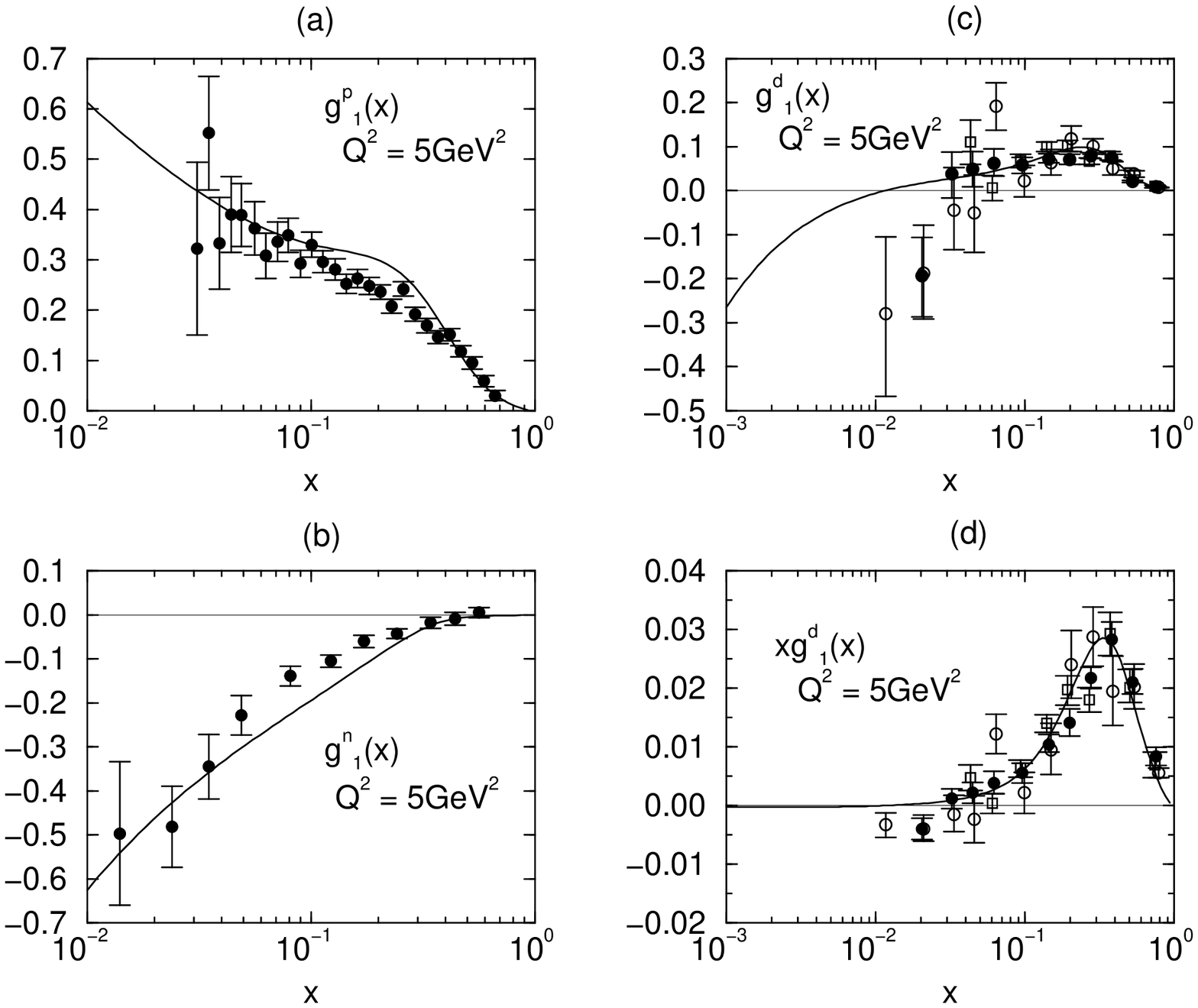}
\end{center}  
\renewcommand{\baselinestretch}{1.00}
\caption{\small{The theoretical predictions for the longitudinally polarized
structure functions for the proton, the neutron and the deuteron
at $Q^2 = 5 \,\mbox{GeV}^2$ in comparison with the corresponding
experimental data. The filled circles in (a) and (b) respectively
corresponds to the E143 and the E154 data,
whereas the filled circles, the open circles and the open squares
in (c) and (d) represent the E143, the E155 and
the SMC data.}}
\renewcommand{\baselinestretch}{1.00}
\end{figure}

Given in Fig.2 is a direct comparison with the NMC data for the
unpolarized nucleon structure functions. One sees that the difference
and the ratio of the proton and neutron structure functions are well
reproduced at least qualitatively except for the ratio at the values
of $x$ close to 1. Integrating the above difference over $x$, we obtain
$S_G = 0.204$ for the so-called Gottfried sum, which is qualitatively
consistent with the NMC analysis, $S_G^{(exp)} = 0.228 \pm 0.007$.

Next, in Fig.3, we compare the predictions for the longitudinally
polarized structure functions of the proton, the neutron and the
deuteron with the corresponding EMC and SMC data.
An excellent feature of the CQSM is a good reproduction of the neutron 
data, which can also be interpreted as a manifestation of chiral symmetry 
in high energy observables \cite{WK99}.
Another interesting observation is that the theory closely follows
the sign change of the recent deuteron data in the small $x$
region \cite{WW00A}. I emphasize that this sign change of the 
theoretical structure function can be traced back to the small $x$
behavior of the previously-discussed isoscalar distribution function
$\Delta u(x) + \Delta d(x)$.
\begin{figure}[th]
\begin{center}
\epsfxsize=7.0cm\leavevmode\epsfbox{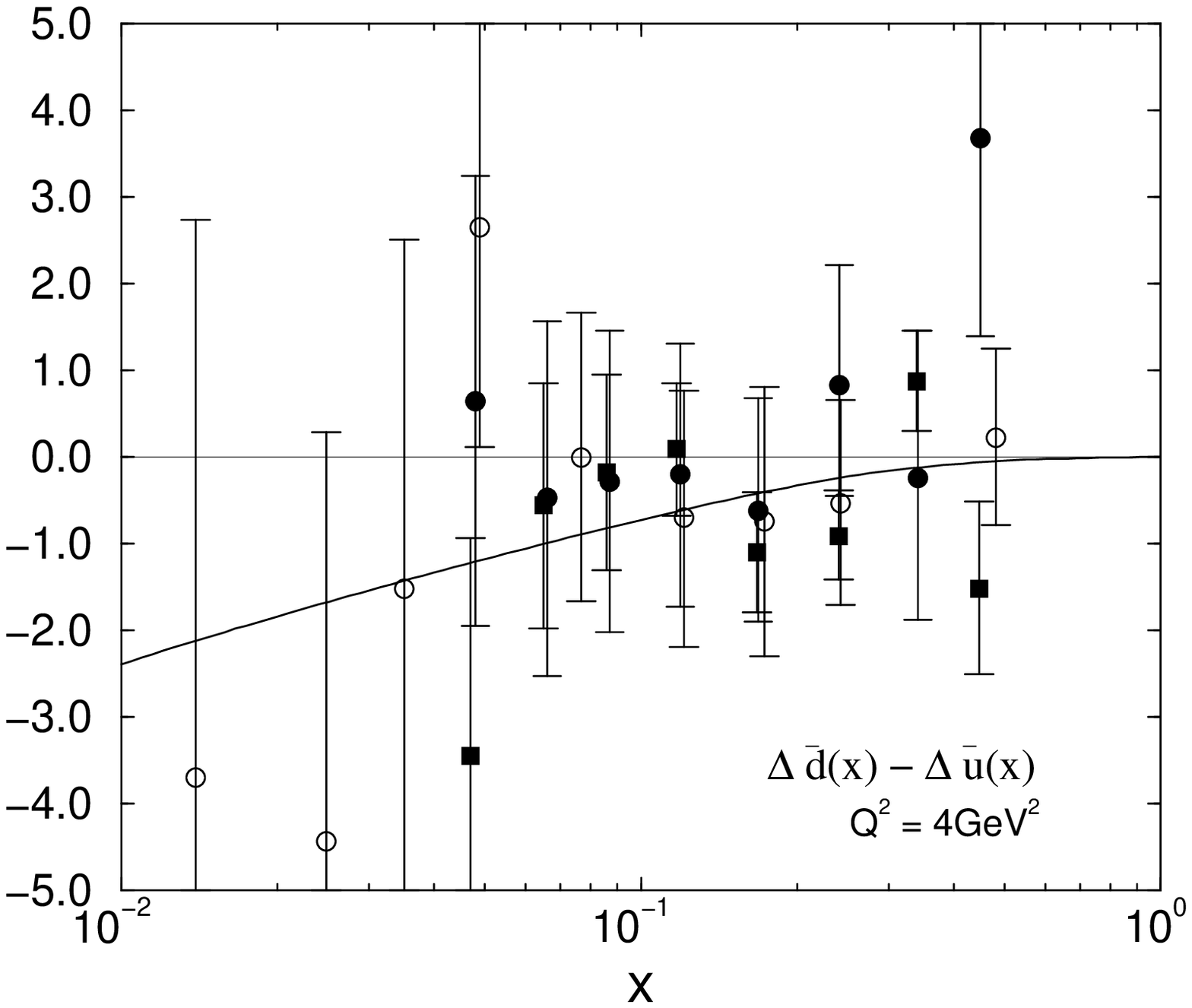}
\end{center}
\renewcommand{\baselinestretch}{1.00}
\caption{\small{The theoretical predictions for the longitudinally polarized
structure functions for the proton, the neutron and the deuteron
at $Q^2 = 5 \,\mbox{GeV}^2$ in comparison with the corresponding
experimental data. The filled circles in (a) and (b) respectively
corresponds to the E143 and the E154 data,
whereas the filled circles, the open circles and the open squares
in (c) and (d) represent the E143, the E155 and
the SMC data.}}
\renewcommand{\baselinestretch}{1.00}
\end{figure}
\begin{figure}[hbtp]
\begin{center}
\epsfxsize=12.0cm\leavevmode\epsfbox{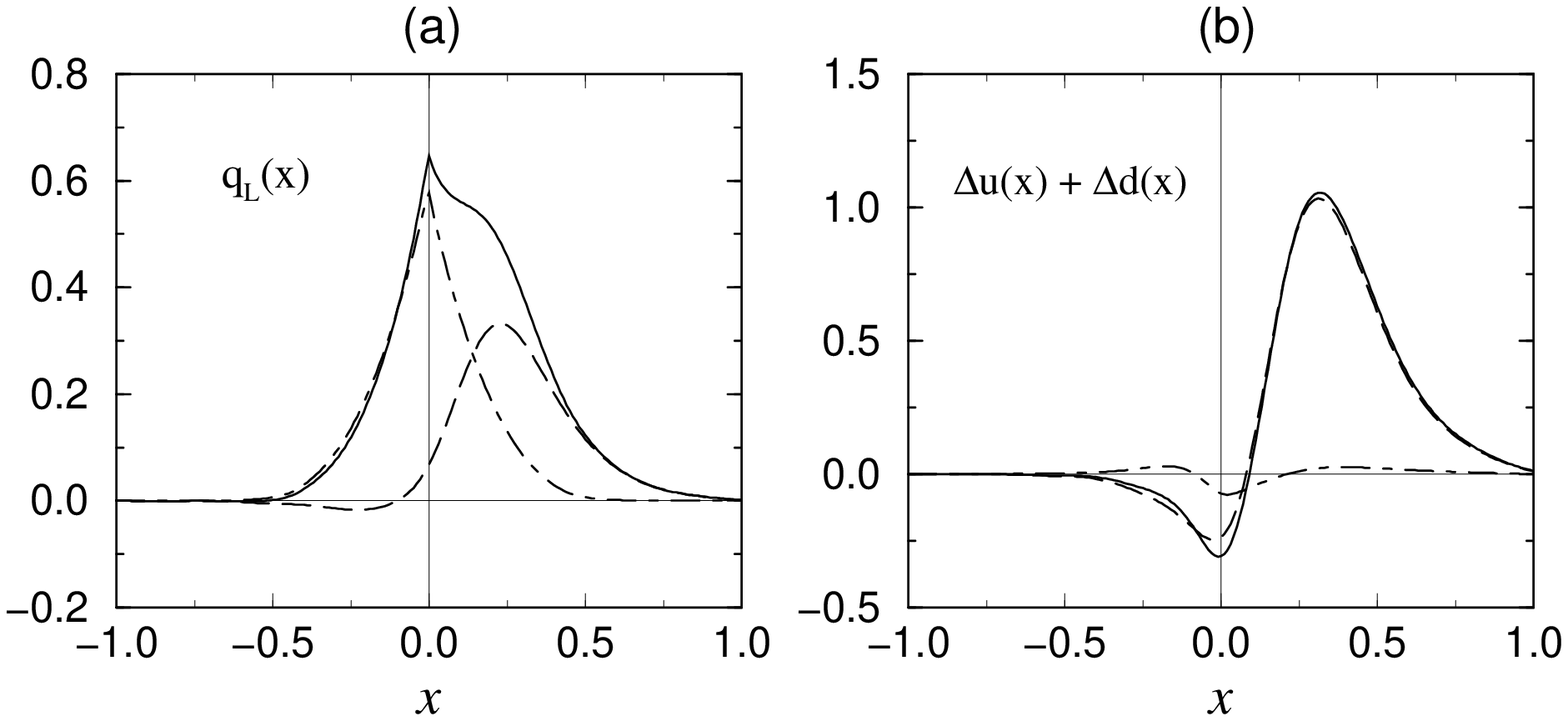}
\end{center}
\renewcommand{\baselinestretch}{1.00}
\caption{\small{(a) The theoretical predictions of the CQSM for the quark
and antiquark orbital angular momentum distribution functions
$q_L(x)$ and (b) the isosinglet quark polarization
$\Delta u(x) + \Delta d(x)$.
The long-dashed and dash-dotted curves respectively stand for
the contributions of the discrete valence level and that of the
Dirac continuum in the self-consistent hedgehog background,
whereas their sums are shown by the solid curves. The distributions
with negative $x$ are to be interpreted as the antiquark
distributions.}} 
\renewcommand{\baselinestretch}{1.00}
\end{figure}

Now we are in a position to discuss a unique prediction of the 
model, i.e. the possible isospin asymmetry of the longitudinally 
polarized sea-quark distribution function.
Very little is known for the spin-dependent sea-quark distributions,
because inclusive data alone cannot give enough constraint to fix them.
Recently, Morii and Yamanishi extracted 
$\Delta \bar{d} (x) - \Delta \bar{u} (x)$ by using polarized
semi-inclusive data, too \cite{MY99}.
Here, the result of their analysis is compared with the 
prediction of the CQSM \cite{WW00A}.
Although the uncertainties of their analysis is 
too large to draw any decisive conclusion, what is still interesting is
their fit given in the following form :
\begin{equation}
 \Delta \bar{d}(x) - \Delta \bar{u}(x) \ = \ C \,x^\alpha \,
 \left( \,\bar{d}(x) - \bar{u}(x) \,\right) \, ,
\end{equation}
with
\begin{equation}
 C \ = \ - \, 3.40, \ \ \ \alpha \ = \ 0.567 \, .
\end{equation}
Here, the negative value of C means that the sign of
$\Delta \bar{d} - \Delta \bar{u}$ is opposite to that of
$\bar{d} - \bar{u}$, which is just consistent 
with the theoretical prediction of the CQSM :
\begin{equation}
 C \ \simeq \ - \,2.0, \ \ \ \ \alpha \ \simeq \ 0.12 \, .
\end{equation}
The fact that C is larger than one is qualitatively consistent with 
the idea of $N_c$-counting mentioned before. Undoubtedly, the underlying
physics behind this $N_c$-counting rule is strong spin-isospin
correlation imbedded in the hedgehog soliton picture.
We emphasize that the Meson Cloud Convolution Model would not predict
large spin polarization of sea quarks, since the pion carries no spin 
and the effect of heavier meson clouds would be much weaker.
\renewcommand{\arraystretch}{1.5}
\vspace{6mm}
\begin{table}[htb]
\caption{\small{The separate contributions of quarks and antiquarks to
the first moment $\Delta \Sigma$ and $L_q$ at the scale of the
model.}}
\begin{center}
\begin{tabular}{cccc} \hline
 & \ \ \ quark \ \ \ & \ \ \ antiquark \ \ \ & \ \ \ total \ \ \ \\
\hline \hline
\ \ \ $\Delta \Sigma$ \ \ \ & 0.40 & - \,0.05 \, & 0.35
 \\ \hline
\ \ \ $2 \,L_q$ \ \ \ & 0.46 & 0.19 & 0.65
 \\ \hline
\ \ \ $\Delta \Sigma \,+ \,2 \,L_q$ \ \ \ &
0.86 & 0.15 & 1.00 \\ \hline
\end{tabular}
\end{center}
\end{table}
\renewcommand{\arraystretch}{1.0}

Also very interesting is another consequence of the chiral soliton
picture of the nucleon. Shown in Fig.5 are the spin and the orbital
angular momentum distribution functions at the model energy
scale \cite{WW00B}.
One notices that the Dirac-sea contribution to the orbital
angular momentum distribution 
function is sizably large and peaked around $x \simeq 0$.
Among others, large support in the negative $x$ region suggests that
sizable amount of orbital angular momentum is carried by
antiquarks. It can also be confirmed from Table 1 for
the corresponding 1st moment \cite{WY91},\cite{WW00B}.
One sees that only $35 \%$ of the total nucleon spin comes from
quark spin, while the remaining $65 \%$ is due to orbital angular
momentum of quarks and antiquarks.

It may be interesting to compare these unique predictions of the CQSM
with the recent lattice QCD study of the nucleon spin
contents \cite{MDLMM99}.
From the analysis of the energy momentum tensor form factor, they
gave an estimate that about $60 \%$ of the total nucleon
spin is carried by quark field. Further combining this result with
the previous estimates of the quark spin content
$\langle \Sigma \rangle$, they
concluded that, out of this $60 \%$, about $25 \%$ comes from
intrinsic quark spin, while the remaining $35 \%$ is attributed
to quark orbital angular momentum.
Since the CQSM at the present level of approximation contains no
explicit gluon fields, let us tentatively renormalize its
prediction by multiplying the above number $60 \%$ of the quark
angular momentum fraction in the lattice simulation, thereby
obtaining the following numbers :
\begin{eqnarray*}
  60 \,\% \ \ \mbox{of} \ \ {\langle \frac{1}{2} \,\Sigma
  \rangle}^{CQSM}
  \ \ &\simeq& \ \ 21 \,\% \, ,\\
  60 \,\% \ \ \mbox{of} \ \ {\langle \,L_q
  \rangle}^{CQSM}
  \ \ \ &\simeq& \ \ 39 \, \% \, .
\end{eqnarray*}
One sees that these numbers are rather close to the corresponding
numbers $25 \,\%$ and $35 \,\%$ in the lattice QCD.
At least, one can confirm an interesting common feature, i.e.
the dominance of the orbital angular momentum part over the
intrinsic spin one.

\begin{figure}[htbp]
\begin{center}
\epsfxsize=13.0cm\leavevmode\epsfbox{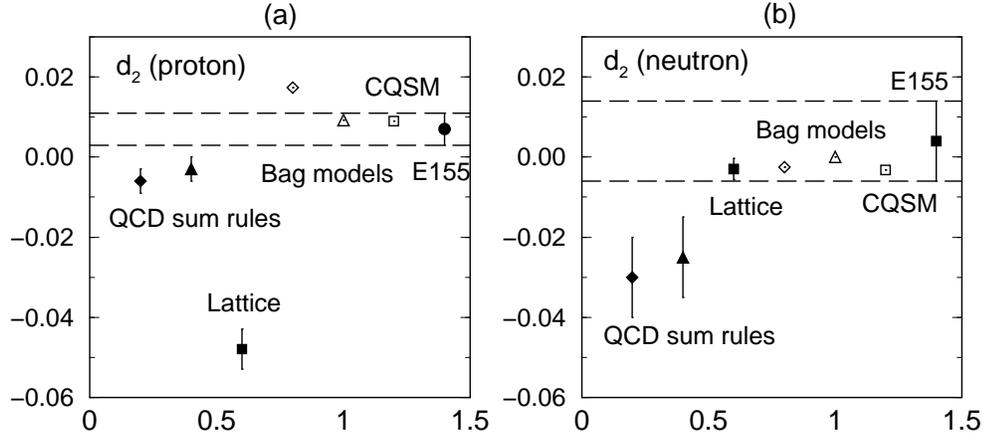}
\end{center}
\renewcommand{\baselinestretch}{1.00}
\caption{\small{The predictions of various theoretical models for the
twist-3 matrix element $d_2$ for the proton and the neutron
are compared with the recent E155 analysis \cite{E155G2}.
Shown theoretical models are from left to right : QCD sum rules
\cite{STEIN95},\cite{BBK90},
lattice QCD \cite{GOCKEL96}, MIT bag models \cite{JU94},\cite{SHT95},
and the CQSM \cite{W00}.}}
\renewcommand{\baselinestretch}{1.00}
\end{figure}

Although qualitatively interesting, one should not put too much
confidence on the precise numbers obtained in the lattice QCD
calculation at the present stage.
To illustrate it, we compare in Fig.6 its predictions for the twist-3
matrix element of $g_2 (x)$ with those of other models including the
CQSM \cite{W00}.
To be more precise, what is compared here is the third moment of
the twist-3 part of the spin structure function $g_2 (x)$.
\begin{equation}
  d_2 (Q^2) \ \ \equiv \ \ 3 \ \int_0^1 \,\,x^2 \,\,
  \bar{g}_2 (x, Q^2) \,\,dx \ \ = \ \
  2 \ \int_0^1 \,\,x^2 \,\,\left[ \,g_1 (x,Q^2) \ + \
  \frac{3}{2} \,\,g_2 (x,Q^2) \,\right] \,\,dx \, .
\end{equation}
The lattice QCD as well as the QCD sum rule fail to reproduce
$d_2$ of the proton and neutron at the same time, while the CQSM
as well as the MIT bag model give relatively good reproduction
of the E155 data \cite{E155G2}.

\begin{figure}[htbp]
\begin{center}
\epsfxsize=12.0cm\leavevmode\epsfbox{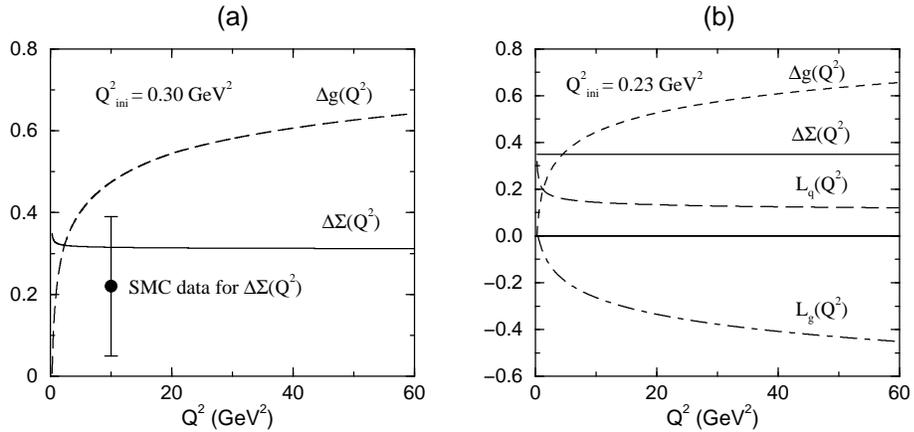}
\end{center}
\renewcommand{\baselinestretch}{1.00}
\caption{\small{(a) NLO evolution of $\Delta \Sigma (Q^2), \Delta g (Q^2)$
\ (b) LO evolution of full nucleon spin contents}}
\renewcommand{\baselinestretch}{1.00}
\end{figure}

The spin and orbital angular momentum contents of the nucleon
are of course scale-dependent quantities. Here, Fig.7(a)
shows the scale dependence of the quark and gluon
longitudinal polarization at the NLO, while Fig.7(b)
does the scale dependence of the full nucleon spin contents
including the orbital angular momentum but at the LO \cite{WW00B}.
The initial energy scale of the LO evolution equation is chosen
to be a little smaller than the NLO case, so that both give
a similar $Q^2$-evolution for $\Delta g$.
At the NLO with $\overline{MS}$ scheme, $\Delta \Sigma$ is
known to have weak scale dependence mainly at low $Q^2$.
The theoretical value $\Delta \Sigma = 0.31$ obtained at
$Q^2 = 10 \,\mbox{GeV}$ is qualitatively consistent with the result of
the recent SMC analysis $\Delta \Sigma^{(exp)}_{SMC} = 0.22 \pm 0.17$
\cite{SMC98}.
Another remarkable feature is that the gluon
polarization $\Delta g$ grows rapidly as $Q^2$ increases,
even if we assume zero polarization at the low renormalization point.

\vspace{6mm}
\noindent
\begin{large}
{\bf 4. Conclusion}
\end{large}
\vspace{3mm}

To sum up, I have shown that, {\it without
introducing any adjustable parameter} except for the
initial-energy scale of
the $Q^2$-evolution, the CQSM can explain all the qualitatively
noticeable features of the recent high-energy deep-inelastic scattering
observables. It naturally explains the excess of $\bar{d}$-sea over
the $\bar{u}$-sea in the proton. It also reproduces qualitative
behavior of the observed longitudinally polarized structure functions
for the proton, neutron and the deuteron.
The most puzzling observation, i.e. unexpectedly small quark spin
fraction of the nucleon can also be explained in no need of
large gluon polarization at the low renormalization point.
Finally, as a unique prediction of the model, I pointed out the
possibility of large isospin asymmetry of the spin-dependent sea
quark distributions, which has also been suggested by the recent
semi-phenomenological analysis of Morii and Yamanishi.
I emphasized that the obtained result $|C| > 1$ in the parametrization
$\Delta \bar{d} (x) - \Delta \bar{u} (x) \,= \,C \,x^{\alpha}
\, ( \bar{d} (x) - \bar{u} (x) )$ is a natural consequence
of the $N_c$-counting rule, but it appears inconsistent with the naive
Meson Cloud Convolution Model. Then, if this large asymmetry of the
longitudinally polarized sea is experimentally established, it
would offer a strong evidence in favor of nontrivial spin-isospin
correlation imbedded in the ``large $N_c$ chiral soliton picture''
of the nucleon.

\vspace{3mm}
The talk is based on the collaborations with T.~Watabe and T.~Kubota.
\vspace{-8mm}

%
%
\vspace{4mm}
\setlength{\baselineskip}{5mm}


\begin{thebibliography}{99}
\bibitem{EMC88} EMC Collaboration, J.~Ashman et al.,
Phys. Lett. {\bf B206} (1988) 364-370 ; \\
Nucl. Phys. {\bf B328} (1989) 1-35.
\bibitem{NMC91} NMC Collaboration, P.~Amaudruz et al.,
Phys. Rev. Lett. {\bf 66}, 2712 (1991).
\bibitem{Kumano98} See, for instance, S.~Kumano,
Phys. Rep. {\bf 303} (1998) 183.
\bibitem{WY91} M.~Wakamatsu and H.~Yoshiki, Nucl. Phys. 
{\bf A524} (1991) 561.
\bibitem{DPP88} D.I.~Diakonov, V.Yu.~Petrov, and P.V.~Pobylitsa,
Nucl. Phys. {\bf B306} (1988) 809.
\bibitem{DPPPW96} D.I.~Diakonov, V.Yu.~Petrov, P.V.~Pobylitsa,
M.V.~Polyakov, and C.~Weiss, \\
Nucl. Phys. {\bf B480} (1996) 341;
{\it ibid.}, Phys. Rev. {\bf D56} (1997) 4069.
\bibitem{WK98} M.~Wakamatsu and T.~Kubota, Phys. Rev. {\bf D57}
(1998) 5755.
\bibitem{WK99} M.~Wakamatsu and T.~Kubota, Phys. Rev. D{\bf 60}
(1999) 034020.
\bibitem{HKM98} M.~Hirai, S.~Kumano, and M.~Miyama, Compt.
Phys. Commun. {\bf 108} (1998) 38; \\
{\it ibid.}, {\bf 111} (1998) 150.
\bibitem{PPGWW99} P.V.~Pobylitsa, M.V.~Polyakov, K.~Goeke, T.~Watabe
and C.~Weiss, \\
Phys. Rev. {\bf D59} (1999) 034024.
\bibitem{WW00A} M.~Wakamatsu and T.~Watabe, Phys. Rev. {\bf D62}
 (2000) 017506.
\bibitem{MY99} T.~Morii and T.~Yamanishi, Phys. Rev. {\bf D61} (2000)
\bibitem{WW00B} M.~Wakamatsu and T.~Watabe, Phys. Rev. {\bf D62}
 (2000) 054009.
\bibitem{W00} M.~Wakamatsu, Phys. Lett. {\bf B487} (2000) 118.
\bibitem{MDLMM99} N.~Mathur, S.J.~Dong, K.F.~Liu, L.~Mankiewics,
and N.C.~Mukhopadhyay, hep-ph / 9912289.
\bibitem{STEIN95}  E.~Stein, Phys. Lett. B343 (1995) 369.
\bibitem{BBK90} I.~Balitsky, V.~Braun and A.~Klesnichenko,
Phys. Lett. B242 (1990) 245 ; \\
B318 (1993) 648(E).
\bibitem{GOCKEL96} M.~G\"{o}ckeler et al., Phys. Rev. D53 (1996) 2317.
\bibitem{JU94} X.~Ji and P.~Unrau, Phys. Lett. {B333} (1994) 228.
\bibitem{SHT95} F.M.~Steffens, H.~Hoffmann and A.W.~Thomas,
Phys. Lett. B358 (1995) 139.
\bibitem{SONG96} X.~Song, Phys. Rev. D54 (1996) 1955.
\bibitem{E155G2} E155 Collaboration, P.L.~Anthony et al.,
Phys. Lett. B458 (1999) 529.
\bibitem{SMC98} SMC Collaboration, B.~Adeva et al., Phys. Rev.
{\bf D58} (1998) 112001.
\end{thebibliography}
\end{document}